\documentclass[a4paper]{article}

\usepackage{a4wide}

\usepackage[english]{babel} 
\usepackage{float}
\usepackage{amssymb,amsmath}
\usepackage{algorithm}
\usepackage{algpseudocode}
\usepackage{capt-of}
\usepackage{graphicx}

\usepackage{xspace}
\usepackage{tabularx}  % To have the table fill out the page

\usepackage{hyperref}
\usepackage{caption}

\usepackage{multirow}

\renewenvironment{abstract}
 {\small
  \begin{center}
  \bfseries \abstractname\vspace{-.5em}\vspace{0pt}
  \end{center}
  \list{}{
    \setlength{\leftmargin}{.5cm}%
    \setlength{\rightmargin}{\leftmargin}%
  }%
  \item\relax}
 {\endlist}

\makeatletter
\renewcommand\fs@ruled{%
  \def\@fs@cfont{\rmfamily}%
  \let\@fs@capt\floatc@plain%
  \def\@fs@pre{\hrule height.8pt depth0pt \kern2pt}%% or use \def\@fs@pre{} to get rid of the top rule
  \def\@fs@post{\kern2pt\hrule\relax}%% or use \def\@fs@post{} to get rid of the last rule
  \def\@fs@mid{\kern2pt\hrule\kern2pt}%% or use \def\@fs@mid{} to get rid of the middle rule
  \def\@fs@post{}
  \let\@fs@iftopcapt\iffalse}
\makeatother

\usepackage{authblk}

\title{Variational Bayesian hierarchical regression for data analysis }
\author[1]{Dennis Becker \thanks{dbecker@leuphana.de}}
\affil[1]{Department of Information Systems, Leuphana University}

\usepackage[T1]{fontenc}

\begin{document}

  \maketitle

\centerline{Paper Draft - arxiv.org }

\begin{abstract}
Collected data, which is used for analysis or prediction tasks, often have a hierarchical structure, for example, data from various people performing the same task. Modeling the data's structure can improve the reliability of the derived results and prediction performance of newly unobserved data. Bayesian modeling provides a tool-kit for designing hierarchical models. However, Markov Chain Monte Carlo methods which are commonly used for parameter estimation are computationally expensive. This often renders its use for many applications not applicable. However, variational Bayesian methods allow to derive an approximation with much less computational effort. This document describes the derivation of a variational approximation for a hierarchical linear Bayesian regression and demonstrates its application to data analysis.
\end{abstract}

\section{Introduction}

Bayesian methods allow to develop models that describe the data generation process, derive confidence bounds on its estimated parameters and predictions of new observations.
The Bayesian inference to derive the model parameters, however, especially for hierarchical models, quickly becomes intractable. But the analyzed data often have a hierarchical structure. For example, in social science data can be observed from different people that perform the same task. Therefore, it appears natural to model such a structure to derive an estimate for each subject, provide individual predictions of future observations, and conclude about the general population.

For the estimation of such models usually, Markov Chain Monte Carlo (MCMC) methods are used, which approximate the model by sampling from a Markov chain with the stationary distribution of the posterior distribution~\cite{hastings70,Geman1984}. Although these methods provide guarantees about the samples that are taken from the targeted density~\cite{Robert:2005:MCS:1051451}, they are computationally expensive even for small data sets. 

Thus, variational inference can be suited for larger data sets and scenarios where the model or a variety of models have to be estimated more quickly.
It, however, only provides an approximation of the posterior distribution, does not provide the same guarantees such as MCMC methods, and underestimates the variance of the posterior distribution~\cite{Jordan:1999:IVM:339248.339252}. 

In the following, we will describe the derivation of a variational Bayesian hierarchical regression model.
The model is very closely related to the  Bayesian linear regression model that has been described by Drugowitsch (2013)~\nocite{Drugowitsch2013} and Bishop (2006)\nocite{Bishop:2006:PRM:1162264}, to which we add a hierarchical prior distribution. 
After deriving the model, we will demonstrate the use on a freely available data set~\cite{GunduzFokoue:2013}.
An implementation of the algorithm in R is available:\url{https://github.com/dennisthemenace2/hBReg/}

\section{Variational Inference}
%%%
Probabilistic models enable to describe how the observed variables $\mathbf{X}= x_i,...,x_n$ have been generated by the influence of a number of latent variables $\mathbf{Z}$. Under these latent variables, we summarize all parameters that are used to model the relations in the observed data. The derived probabilistic model is specified by the joint density $p(\mathbf{X}, \mathbf{Z})$. To derive an estimate of the latent variables given the observed data we aim to compute $p(\mathbf{Z} | \mathbf{X})$ which is called the posterior distribution. Using Bayes' rule the posterior distribution is stated as:
\begin{equation*}
p(\mathbf{Z} | \mathbf{X}) =  \frac{p(\mathbf{X},\mathbf{Z}) }{p(\mathbf{X})} =\frac{p(\mathbf{X} | \mathbf{Z}) p(\mathbf{Z})}{p(\mathbf{X})}= \frac{p(\mathbf{X} | \mathbf{Z}) p(\mathbf{Z})}{\int p(\mathbf{X}, \mathbf{Z}) d\mathbf{Z}}.
\end{equation*}
The denominator is called the marginal likelihood or model evidence.
Typically, the calculation of the evidence is unavailable in closed-form, therefore inference in such models is often based on Markov Chain Monte-Carlo methods~\cite{hastings70,Geman1984}. 
Alternatives are approximative methods such as variational inference~\cite{Beal2003,Jordan:1999:IVM:339248.339252}. 

Variational inference aims to approximate the posterior distribution $p$ with a distribution $q$ from a set of tractable distribution $Q$. It, therefore, states the inference problems as an optimization problem, which allows to estimate an approximate solution $ q(\mathbf{Z}) \approx p(\mathbf{Z} | \mathbf{X})  $.

To solve this optimization problem, we need a measure of the similarity between $q$ and $p$. The most common type of variational Bayes to describe the difference between these two distributions is the Kullback-Leibler ($KL$)\cite{Kullback51klDivergence} divergence which measures the differences in the information contained within two distributions:
\begin{equation}
  KL(q||p) = \sum_x q(x) \log \frac{q(x)}{p(x)}.
\label{equ:kl}
\end{equation}
The difference between both distributions is always greater or equal to zero $KL(q||p)\geq 0$ for all  $q$,$p$, and only equal to zero if $​q=p$. Next, we plug the posterior distribution $p(\mathbf{Z} | \mathbf{X})$ into the $KL$ divergence:

\begin{align*}
KL(q(\mathbf{Z})||p(\mathbf{Z}|\mathbf{X}))
& = \sum_\mathbf{Z} q(\mathbf{Z}) \log \frac{q(\mathbf{Z})}{p(\mathbf{Z}|\mathbf{X})} \\
& = \sum_\mathbf{Z} q(\mathbf{Z}) \log \frac{q(\mathbf{Z})}{p(\mathbf{X},\mathbf{Z})} + \log p(\mathbf{X}) \\
& = KL(q(\mathbf{Z})||p(\mathbf{X},\mathbf{Z})) + \log p(\mathbf{X}) 
\end{align*}

First, we pulled the normalizing constant, or marginal probability $p(\mathbf{X})$ out, and recognize the remaining term as the $KL$ divergence of the unnormalized
posterior distribution. By maximizing the unnormalized $KL$ divergence, we minimize the above defined $KL$ divergence. With respect to the variational distribution $q(\mathbf{Z})$, the log model evidence $\log p(\mathbf{X})$ is constant. Since $KL(q(\mathbf{Z})||p(\mathbf{Z}|\mathbf{X}))\geq 0$, we get by rearranging terms that
\begin{equation*}
  \log p(\mathbf{X})  = KL(q(\mathbf{Z})||p(\mathbf{Z}|\mathbf{X})) -KL(q(\mathbf{Z})||p(\mathbf{X},\mathbf{Z})).
\end{equation*}
We notice that the log model evidence $\log p(\mathbf{X})$ is equal to the difference between the normalized and unnormalized $KL$ divergence.
Furthermore, the $KL$ divergence of the unnormalized posterior distribution is a lower bound on the model evidence.  Due to this property, the negative $KL(q(\mathbf{Z})||p(\mathbf{X},\mathbf{Z}))$ is called the variational lower bound or the evidence lower bound (ELBO). Minimizing the lower bound amounts to maximizing a lower bound on the model evidence.

To complete the specification of the optimization problem, we need to describe the variational family $Q$.
A widely used class of distributions is the mean-field approximation, which assumes an independent factorization:
\begin{equation*}
q(\mathbf{Z}) = \prod_{m=1}^{M} q_{m}(Z_{m}).
\end{equation*}

Each latent variables $Z_{m}$ can be governed by its own variational factor $q_{m}(Z_{m})$, which renders the individual factors mutually independent.
For the mean-field choice of $Q$, we can optimize the problem using coordinate descent. We iterate over the variational factors $q_{m}(Z_{m})$ and for each $m$ we optimize the evidence lower bound over $q_{m}$ while keeping the other variational factors $q_{j}(Z_{j})$ constant. This results into the log of the optimal solution for each factor:
\begin{equation}
\ln q_{m}(Z_{m}) = \mathbb{E}_{j \neq m}\Big[\ln p(\mathbf{X}, \mathbf{Z})\Big] + \text{const}.
\label{equ:corddesc}
\end{equation}

This procedure iteratively fits the fully-factored $q(\mathbf{Z}) = q_1(Z_1) q_2(Z_2) \cdots q_m(Z_m)$ approximation of $p(\mathbf{Z}|\mathbf{X})$.

\subsection{Hierarchical Variational Bayesian Regression }

An often encountered scenario is that the analyzed data has a hierarchical structure, for example, people are undergoing the same treatment type.
Then we can consider the data from each client as an independent sample from the same population, which naturally suggest a hierarchical structure of the data.
Typically, the observed data consist of some target variable $\mathbf{y}_i\in \mathbb{R}$ and independent variables $\mathbf{x}_i \in \mathbb{R}^{D}$ for each client. If we assume a linear relationship between those, the suggested model is a hierarchical linear regression.

%% $\mathbf{D}={\mathbf{Y_i},\mathbf{X_{id}}}$  

\begin{figure}[H]%
    \centering
   \includegraphics[width=0.6\textwidth]{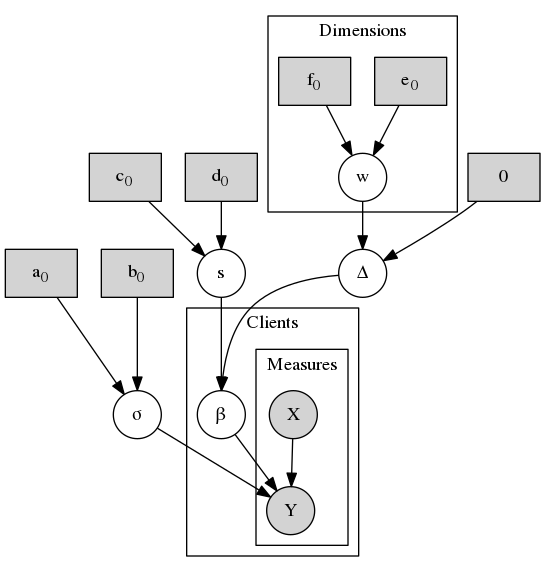}
    \caption{Plate notation of the hierarchical linear regression model}%
    \label{fig:plate}%
\end{figure}

We further assume that we want the benefits of a Bayesian model and that it has to be estimated quickly, therefore we want to derive a hierarchical Bayesian linear regression model. The derived model is very similar to the Bayesian linear regression model derived by Drugowitsch (2013)\nocite{Drugowitsch2013} and Bishop (2006)\nocite{Bishop:2006:PRM:1162264}. 
In addition to these models, we have the hierarchical prior for the individual people or clients' weights. The complete model is shown in plate notation in Figure~\ref{fig:plate}.

The model is shown with all components, which makes it appear quite confusing, but this allows to talk about the design of the probabilistic graphical model and about the reasoning behind it.
The filled nodes are observed or known and circles typically represent distributions, whereas boxes represent constants. The rectangles with names in it are plates, which indicate that their elements are observed mutable times.
We start from the bottom of the illustration with the observed data $\mathbf{Y}$, regarding the nature of the target variable we assume a continues variable for the regression model.
This suggests a normal distribution $\mathcal{N}(y\mid \mu,\sigma)$, which has two parameters, the mean value $\mu$ and a variance parameter $\sigma$. These parameters have both to be specified by either a distribution or a constant value. That is why these models naturally appear to grow in one direction.
Next, we discuss the plates surrounding the data. These illustrate that we have multiple clients or subjects from which we observed repeated measures or observations.
We specify the inverse of the variance parameter using the conjugate prior a gamma distribution $\operatorname{Gamma}(\sigma|a_0,b_0)$. The Gamma distribution has two parameters that we need to specify. For these values, we assume both small values, which represents an uninformative prior. The specification of this prior distribution on the precision leads to an estimation of the noise from the data. We continue with specifying the $\mu$ parameter of the normal distribution which is the target variable. Based on the linear relationship between the observed data and the target variable we use matrix multiplication of the observed data $\mathbf{x}_{ci}$ and a client-specific weight vector $\beta_c$ which is also represented by the conjugate prior of the mean value a normal distribution. This completes the description of the observed data and we sate the factors:
\begin{align*}
  p(\mathbf{Y}\mid \mathbf{x}\beta_i,\sigma) & = \prod_{i=1}^C \prod_{c=1}^M \mathcal{N}(y_{c,m}\mid \mathbf{x}_{c,m}\beta_i,\sigma^{-1}), \\
  p(\sigma) & = \operatorname{Gamma}(\sigma\mid a_0, b_0).\\
\end{align*}
In an iterative fashion, we continue describing the client individual parameters $\beta_i$. The prior on the mean value of the client individual parameters which is titled $\Delta$ represents the population of which all individuals are samples. The precision $s$ describes the influence of population or hierarchical prior on the individual. We describe the clients specific factors as:
\begin{align*}
 p(\beta_i\mid \Delta,s) & = \prod_{i=1}^C  \mathcal{N} \left (\beta_i \mid \Delta, s^{-1} \right ) \\
  p(s) & = \operatorname{Gamma}(s\mid c_0, d_0)\\
\end{align*}

For the prior on the mean of the hierarchical prior $\Delta$, we use the constant 0, which encourages the weights to become small. It shrinks the weights towards 0 similar to a quadratic regularisation term in ridge regression~\cite{Bishop:2006:PRM:1162264}. For the prior on the precision, we chose a Gamma distribution for each observed variable (or Dimension) individually. The reasoning is that during optimization of the model irrelevant coefficients will shrink automatically. This process is known as automatic relevance determination (ARD)\cite{MacKay1992,Tipping2000,Wipf2008}.
This concludes the model construction with the final factors:
\begin{align*}
  p(\Delta\mid 0,\mathbf{w}) & =  \mathcal{N}(\Delta \mid 0,\ \mathbf{w}^{-1}) \\
  p(\mathbf{w}) & =  \prod_{d=1}^D \operatorname{Gamma}(\mathbf{w}_d \mid e_0, f_0)\\
\end{align*}

%%% model derivation

In order to continue with the model development and variational approximation, we state the models joint probability:

\begin{align*} 
  p(\mathbf{Y},\mathbf{x},\beta_i,\sigma,\Delta,s,\mathbf{w}) &= p(\mathbf{Y}\mid \mathbf{x}\beta_i,\sigma) p(\beta_i \mid \Delta,s) p(\Delta\mid 0,w) p(\sigma) p(s) p(\mathbf{w}),\\
\end{align*}

and assume that the posterior distribution is approximated by the factored variational posterior distributions,
\begin{align*}
  p(\beta_i,\Delta, \sigma,s,w \mid \mathbf{D} )\approx q(\beta_i)q(\Delta)q(\sigma)q(s)q(\mathbf{w}). 
\end{align*}

Now, we have to derive the update equations for the variational factors using the coordinate descent algorithm by applying Equation~\ref{equ:corddesc}.

\subsection{Update factor q($\beta_i$)}

To derive the update equations, we select the factors that depend on beta, since the others are constant and absorbed into a constant term.
We replace the factors with their actual distributions and multiply everything out to separate the terms that not depend on $\beta, \Delta$ or $\sigma$ out into the constant term.
The terms are rearranged and grouped until we receive a distribution that we recognize. Since we used the conjugate prior and the product of two normal distributions, the result will also be a normal distribution.

\begin{align*}
\ln q_{\beta_i}^{*}(\beta_i) &= \operatorname{E}_{\Delta,\sigma,s}\left[\ln p(\mathbf{y}\mid \mathbf{x}\beta_i,\sigma) + \ln p(\beta_i \mid \Delta,s) \right] + \text{const} \\
&= \operatorname{E}_{\sigma}\left[\sum_{c=1}^M \frac{1}{2}\ln\sigma -\frac{\sigma}{2}(y_{c}-\mathbf{x}_c\beta_i)^2\right]+\operatorname{E}_{\Delta,s}\left[ \frac{1}{2}\ln s- \frac{s}{2}(\Delta -\beta_i)^2 \right]+C \\
&= \operatorname{E}_{\sigma}\left[\sum_{c=1}^M -\frac{\sigma}{2}(y_{c}-x_c\beta_i)^2\right]+\operatorname{E}_{\Delta,s}\left[-\frac{s}{2}(\Delta -\beta_i)^2 \right]+\text{const} \\
&= \operatorname{E}_{\sigma}\left[ -\frac{\sigma}{2}\sum_{c=1}^M{y_c^2} +\sigma\beta_i\sum_{c=1}^M{y_c \mathbf{x}_c} - \frac{\sigma \beta_i^2 \sum_{c=1}^M \mathbf{x}^2_c }{2}        \right]+\operatorname{E}_{\Delta,s}\left[  -\frac{s\Delta^2}{2}+ s\Delta\beta_i - \frac{s\beta_i^2}{2}  \right]+\text{const}\\
&= \operatorname{E}_{\sigma}\left[ \sigma\beta_i\sum_{c=1}^M{y_c \mathbf{x}_c}- \frac{\sigma \beta_i^2 \sum_{c=1}^M \mathbf{x}^2_c }{2} \right]+\operatorname{E}_{\Delta,s}\left[  s\Delta\beta_i - \frac{s\beta_i^2}{2}  \right]+\text{const} \\
&=  \underbrace{(\operatorname{E}_{\sigma}\left[\sigma \right] \sum_{c=1}^M{\mathbf{x}_c y_c}+ \operatorname{E}_{s}\left[ s \right] \operatorname{E}_{\Delta}\left[\Delta \right] )}_{\text{Mean}}\beta_i - \frac{1}{2}\underbrace{( \operatorname{E}_{\sigma} \left[\sigma\right] \sum_{c=1}^M \mathbf{x}^2_c  +\operatorname{E}_{s}\left[s\right])}_{\text{Variance }}\beta_i^2 +\text{const} \\
\end{align*}

By completing the square over $\beta$, we can derive the parameters of the Gaussian distribution.
There are still the expectations of the variables with respect to their variational distribution, which we consider to be constant, that we have to replace.
The variational distribution of the terms $\operatorname{E}_{\sigma} \left[\sigma\right]$ and $\operatorname{E}_{s} \left[s\right]$ is a Gamma distribution and the expected value for a Gamma distribution is $\operatorname{E}\left[\operatorname{Gamma}(a,b)\right] = \frac{a}{b}$. Therefore, the expect values are $\operatorname{E}_{\sigma} \left[\sigma\right] = \frac{a_n}{b_n}$
and $\operatorname{E}_{s} \left[s\right] = \frac{c_n}{d_n}$ respectively. The expectation for $\Delta$ is with respect to a normal distribution, the expected value of a normal distribution is the mean, which results into the expected values of $\operatorname{E}_{\Delta} \left[\Delta\right] =\Delta$. This leads to the following update equations for $\beta_i$:

\begin{align*}
\lambda_{\beta_i} &= \frac{a_n}{b_n}  \sum_{c=1}^M \mathbf{x}^2_c +\frac{c_n}{d_n},\\
\beta_i &=  \frac{\frac{a_n}{b_n}  \sum_{c}^M{\mathbf{x}_c y_c} + \frac{c_n}{d_n} \Delta }{\lambda_{\beta_i}}.
\end{align*}

The update equations shows that the client individual weights are dependent on their individual data and the hierarchical prior.

\subsection{Update factor q($\Delta$)}
Next, we estimate the update equations for the factor q($\Delta$) to derive the variational posterior distribution. 
The procedure is the same as before, we select the terms that depend on $\Delta$ because the other terms get absorbed into the constant term.
We multiply everything out and rearrange until we derive the same form as for the previous factor, where we recognize the terms of the resulting normal distribution.

\begin{align*}
\ln q_{\Delta}^{*}(\Delta) &= \operatorname{E}_{\beta_i,s}\left[\ln p(\beta_i\mid \Delta,s) + \ln p(\Delta \mid \mu_0,\mathbf{w}) \right] + \text{const} \\
&= \operatorname{E}_{\beta_i,s}\left[\sum_{i=1}^C \frac{1}{2}\ln s -\frac{s}{2}(\Delta-\beta_i)^2\right]+\operatorname{E}_{\mathbf{w}}\left[ \frac{1}{2}\ln \mathbf{w} - \frac{\mathbf{w}}{2}(\Delta -\mu_0)^2 \right]+\text{const} \\
&= \operatorname{E}_{\beta_i,s}\left[\sum_{i=1}^C -\frac{s}{2}(\Delta-\beta_i)^2\right]+\operatorname{E}_{\mathbf{w}}\left[ - \frac{\mathbf{w}}{2}(\Delta -\mu_0)^2 \right]+\text{const} \\
&= \operatorname{E}_{\beta_i,s}\left[ -\frac{C s\Delta^2}{2} + s\Delta\sum_{i=1}^{C}{\beta_i} \right]+\operatorname{E}_{\mathbf{w}}\left[ \mathbf{w}\Delta-\frac{w\Delta^2}{2}  \right]+\text{const} \\
&= \operatorname{E}_{\beta_i,s}\left[ -\frac{1}{2}(Cs+w)\Delta^2 \right]+\operatorname{E}_{\mathbf{w},s,\beta_i}\left[ \Delta (s\sum_{i}^{C}{\beta_i} +\mathbf{w}\mu_0)   \right]+\text{const} \\
\end{align*}

Similar to before, we have to fill in the expected values $\operatorname{E}_{\mathbf{w}} \left[\mathbf{w}\right] = \frac{e_n}{\mathbf{f}_n}$, and $\operatorname{E}_{\beta_i} \left[\beta_i\right] = \beta_i$, and the expected value of $s$ has already been stated in the update of the previous factor. This leads to the following update equations for $\Delta$:

\begin{align*}
\lambda_{\Delta}&= C \frac{c_n}{d_n} + \frac{e_n}{\mathbf{f}_n}, \\
\Delta &= \frac{\frac{c_n}{d_n} \sum_{i}^I{\beta_i} +\frac{e_n}{\mathbf{f}_n} \mu_0 }{\lambda_{\Delta} }.
\end{align*}

We indicated that $\mathbf{f_n}$ is a vector with dimensions entries whereas $e_n$ is a single value. Because each weight will have its own Gamma prior, but the parameter $e$ will be the same for all $\mathbf{f_n}$ assuming they have the same prior value. We will see this in the derivation of the update for $\mathbf{w}$. The update equations show that the hierarchical prior is similar to a mean weight, where it consists of the sum of all weights which are scaled by the individual Gamma distributions of $\mathbf{w}$.

\subsection{Update factor  q($\sigma$)}

Now, we derive the update equation for the precision of the noise $\sigma$. We proceed as in the previous updates and since we use a conjugate prior, we expect a Gamma distribution as the result.

\begin{align*}
\ln q_{\sigma}^{*}(\sigma) &= \operatorname{E}_{\beta_i}\left[\ln p(\mathbf{Y}\mid \mathbf{x}\beta_i,\sigma) + \ln p(\sigma \mid a,b) \right] + \text{const} \\
&= \operatorname{E}_{\beta_i}\left[\sum_{i=1}^C \sum_{c=1}^M \frac{1}{2}\ln \sigma -\frac{\sigma}{2}(y_{i,c}-\mathbf{x}\beta_i)^2\right]+\left[ (a-1)\ln \sigma -b\sigma \right] +\text{const}\\
&= \operatorname{E}_{\beta_i}\left[ -\frac{\sigma}{2} \sum_{i=1}^C \sum_{c=1}^{M}{(y_{i,c}-\mathbf{x}\beta_i)^2}\right]+ (a-1)\ln \sigma -b\sigma +\frac{\sum_{i=1}^C \sum_{c=1}^{M}{1}}{2}\ln \sigma +\text{const}\\
&= \operatorname{E}_{\beta_i}\left[ -\frac{\sigma}{2} \sum_{i=1}^C \sum_{c=1}^{M}{y_{i,c}^2 -2y_{i,c}\mathbf{x}\beta_i + \mathbf{x}^2\beta_i^2 }\right]+ ( (a-1)+\frac{\sum_{i=1}^C \sum_{c=1}^{M}{1}}{2} )\ln \sigma -b\sigma +\text{const}\\
&= \operatorname{E}_{\beta_i}\left[ - \underbrace{ ( (\sum_{i=1}^{C}\mathbf{x}\beta_i^2- 2\sum_{i=1}^C \sum_{c=1}^{M}y_{i,c}\mathbf{x}^2\beta_i+\sum_{i=1}^C \sum_{c=1}^{M}{y_{i,c}^2)\frac{1}{2} +b ) }}_{\text{Rate parameter b}} \sigma \right]+ \underbrace{( (a-1)+\frac{\sum_{i=1}^C \sum_{c=1}^{M}{1}}{2} )}_{\text{Shape parameter a}}\ln \sigma  +\text{const}\\
\end{align*}

We recognize the result as a $\operatorname{log} \operatorname{Gamma}$ distribution. This time we need to estimate the expected value of the $\beta^2$, note that $\operatorname{E} \left[\mathbf{X}^2\right] = \operatorname{E} \left[\mathbf{X}\right]+ Var(\mathbf{X})$. This leads to $\operatorname{E}_{\beta_i} \left[\beta_i^2\right] = \beta_i+ \lambda_{\beta_i}$ for the expectation of $\beta_i^2$. If we substitute the expectations and rearrange the terms, we receive the following update equations:

\begin{align*}
N   &= \sum_{i=1}^C \sum_{c=1}^{M}{1},\\
a_n &= a + \frac{N}{2},\\
b_n &= b + \frac{1}{2} \left( \sum_{i=1}^{C} \sum_{c=1}^{M}{(y_{i,c}-\mathbf{x}_{i,c}\beta_i)^2} + \sum_{i}^{C} \mathbf{x}^T_i \lambda_{\beta_i} \mathbf{x}_i  \right).
\end{align*}

Considering the update equations, we notice that it represents the noise of measuring the target with the prediction error term and the second term is the sum of the standard errors.
If we assume that the fitting error on the data is high compared to the number of the sample, the expected value $\frac{a_n}{b_n}$ of the precision would be low, so the variance is high.
Of course, the opposite would be true too if we make no prediction error but have high uncertainty in the estimated clients' weights, we expect the measures to be less reliable.

\subsection{Update factor q(s)}

The result for factor $s$ which regulates the influence of the hierarchical prior on the client individual weights will also be a Gamma distribution.

\begin{align*}
\ln q_{s}^{*}(s) &= \operatorname{E}_{\beta_i,\Delta}\left[\ln p(\beta_i\mid \Delta,s) + \ln p(s \mid c,d) \right] + \text{const} \\
&= \operatorname{E}_{\beta_i,\Delta}\left[\sum_{i=1}^C \sum_{d=1}^D \frac{1}{2}\ln s -\frac{s}{2}(\beta_{i,d}-\Delta_d)^2\right]+\left[(c-1)\ln s -d s \right] +\text{const}\\
&= \operatorname{E}_{\beta_i,\Delta}\left[ \sum_{i=1}^C \sum_{d=1}^D -\frac{s}{2}(\beta_{i,d}-\Delta_d)^2\right]+( (c-1)+\frac{\sum_{i=1}^C \sum_{d=1}^{D}{1}}{2} )\ln s -d s +\text{const}\\
&= \operatorname{E}_{\beta_i,\Delta}\left[ -((\sum_{i=1}^{C}\sum_{d=1}^D \beta_{i,d}^2- 2\sum_{i=1}^C\sum_{d=1}^{D}\beta_{i,d}\Delta_{d}+ \sum_{d=1}^{D}{\Delta_{d}^2} )\frac{1}{2} +d )s \right]+( (c-1)+\frac{CD}{2} )\ln s +\text{const}\\
\end{align*}

The result contains the squared expected values for $\beta$ and $\Delta$. The expected value for $\Delta$ is derived similar to the one of $\beta$ in the previous derivation which results in 
$\operatorname{E}_{\Delta} \left[\Delta^2\right] = \Delta+ \lambda_{\Delta}$. After substituting and rearranging the terms we derive the update equations:

\begin{align*}
c_n &= c + \frac{DC}{2},\\
d_n &= d + \frac{1}{2} \left( \sum_{i=1}^{C} \sum_{d=1}^{D}{  (\beta_{i,d} -\Delta_d)^2} +  Spur(\lambda_\Delta)+  \sum_{i}^{C} Spur(\lambda_{\beta_i}) \right).
\end{align*}

The update equations show that the variance depends on the squared differences between the individual weights and the hierarchical prior, the variance of the clients' data, and the variance of the estimated weights. 
Similarly, if the difference between the hierarchical prior and the clients' weights is large, the hierarchical prior will have less influence on the individual clients' weights. However, if we imagine this term to be zero, then the variance of the clients' data and variance of the estimated weights still contributes. This means, if there is a high variance in the clients' data or variance between the estimated client individual weights, the influence of that prior will be reduced.

\subsection{Update factor $q(\mathbf{w})$}

Finally, we derive the update equation for $\mathbf{w}$. For simplicity, we show the derivation for one of the factors. 
\begin{align*}
\ln q_{\mathbf{w}_d}^{*}(\mathbf{w}_d) &= \operatorname{E}_{\Delta}\left[\ln p(\Delta_d \mid 0,\mathbf{w}_d) + \ln p(\mathbf{w}_d \mid e,f) \right] + \text{const} \\
&= \operatorname{E}_{\Delta}\left[ \frac{1}{2}\ln \mathbf{w}_d -\frac{w}{2}(\Delta_d)^2\right]+\left[(e-1)\ln \mathbf{w}_d -f \mathbf{w}_d \right] +\text{const}\\
&= \operatorname{E}_{\Delta}\left[  -\frac{\mathbf{w}_d}{2}(\Delta_d)^2\right]+( (e-1)+\frac{1}{2} )\ln \mathbf{w}_d  -f \mathbf{w}_d +\text{const}\\
&= \operatorname{E}_{\Delta}\left[ -\mathbf{w}_d(\frac{1}{2}  (-\Delta_d)^2 +f ) \right]+( (e-1)+\frac{1}{2} )\ln \mathbf{w}_d  +\text{const}\\
\end{align*}

The result shows that this is too a log Gamma distribution, where we have to substitute the expected values by the momentums of the distribution. To keep the notation simple, we show the update equation for $\mathbf{f}$ as a vector. 

\begin{align*}
e_n &= e + \frac{1}{2}\\
\mathbf{f}_n &=  f + \frac{1}{2}(\Delta^{2} + trace(\lambda_\Delta))
\end{align*}

If we inspect the update equation for $\mathbf{w}$, we notice how it penalizes the weights for each dimension. 
The first term is a quadratic penalty for being away from zero and the second term adds a penalty which describes the variance in the individual weights. If a particular weight varies among the clients, this is stronger penalized than when it shows less variation among the clients.

\subsection{Variational lower bound}

After we derived all update equations, we know that an iterative updating of the variational factors will maximize the variational lower bound. In order to determine when the algorithm is converged, we estimate the value of the variational lower bound. This allows us to keep track of the optimization process and provides an approximation of the evidence $\ln p(\mathbf{x})$.
Therefore, we plug our model joint distribution and variational distribution into the negative $KL$ divergence from Equation~\ref{equ:kl}. This time, however, we use the continues definition and because the evidence lower bound was defined as the negative divergence, the factors in the log fraction are switched. The variational lower bound $\mathcal{L}(q)$ is then given by: 

\begin{align*}
\mathcal{L}(q) & = \int\int\int\int\int q(\mathbf{\beta},\Delta,\sigma,s,\mathbf{w}) \ln\left\{\frac{  p(\mathbf{\beta},\Delta, \sigma,s,\mathbf{w} \mid \mathbf{D}) }{q(\beta,\Delta,\sigma,s,\mathbf{w})}\right\}d\mathbf{\beta}\;d \Delta\;d \sigma\; d s\; d \mathbf{w} \\
  & = \;\mathbb{E}_{\mathbf{\beta},\Delta,\sigma,s,\mathbf{w}}\Big[\ln  p(\mathbf{\beta},\Delta, \sigma,s,\mathbf{w} \mid \mathbf{D} )\Big] - \mathbb{E}_{\beta,\Delta,\sigma,s,\mathbf{w}}\Big[ q(\beta,\Delta,\sigma,s,\mathbf{w}) \Big] \\
  & = \;\mathbb{E}_{\mathbf{\beta},\sigma}\Big[\ln   p(\mathbf{Y}\mid \mathbf{x}\beta,\sigma) \Big] + \mathbb{E}_{\beta,\Delta,s}\Big[\ln   p(\beta \mid \Delta,s) \Big] + \mathbb{E}_{\mathbf{w}}\Big[\ln   p(\Delta\mid 0,\mathbf{w})  \Big] +   \mathbb{E}_{\sigma}\Big[\ln   p(\sigma)   \Big] \\
  & \quad + \mathbb{E}_{s}\Big[\ln   p(s)   \Big]   + \mathbb{E}_{\mathbf{w}}\Big[\ln  p(\mathbf{w})  \Big] - \mathbb{E}_{\beta}\Big[\ln q(\beta)\Big] - \mathbb{E}_{\Delta}\Big[\ln q(\Delta)\Big] - \mathbb{E}_{\sigma}\Big[\ln q(\sigma)\Big] - \mathbb{E}_{\mathbf{w}}\Big[\ln q(\mathbf{w})\Big]
\end{align*}

The terms that involve expectations of the variational distributions $\operatorname{log} q(\cdot)$ are the entropies $\mathbb{H}(\cdot)$ of that distribution. The various terms are given in the following:

\begin{align*}
 \mathbb{E}_{\mathbf{\beta},\sigma}\Big[\ln   p(\mathbf{Y}\mid \mathbf{x}\beta,\sigma) \Big]& =\frac{N}{2} ( \psi(an) - \ln bn ) - \frac{\sigma}{2}\left( \sum^C (\mathbf{x}_i \beta_i - \mathbf{y}_i)^T(\mathbf{x}_i \beta_i - \mathbf{y}_i)   + \sum^C \mathbf{x}^T_i \lambda_{\beta_i} \mathbf{x}_i \right),  \\
 \mathbb{E}_{\mathbf{\beta},\Delta,\sigma}\Big[\ln   p(\beta \mid \Delta,s) \Big]& =  \frac{C}{2}(\psi(c_n)-\ln d_n )  - \frac{s}{2}\left( \sum^C (\beta_i-\Delta)^T(\beta_i-\Delta) + Spur(\lambda_\Delta) +  \sum^C Spur(\lambda_{\beta_i}) \right),\\ 
\mathbb{E}_{\Delta,\sigma}\Big[\ln  p(\Delta\mid 0,\mathbf{w}) \Big]& = \sum^D \frac{D}{2}\psi(e_n) - \ln f_n - \sum^D  \frac{\mathbf{w}_d}{2} ( \Delta^2_d +  \lambda_{\Delta_{d,d}} ), \\
 \mathbb{E}_{\sigma}\Big[\ln  p(\sigma)  \Big]& = (a_0-1)( \psi(a_n) - \ln bn ) - b_0\sigma ,\\
 \mathbb{E}_{s}\Big[\ln  p(s)  \Big]& = (c_0-1)( \psi(c_n) - \ln d_n) - d_0s ,\\
 \mathbb{E}_{\mathbf{w}}\Big[\ln  p(\mathbf{w})  \Big]& = \sum^D (e_0-1)( \psi(e_{n}) - \ln \mathbf{f}_{n_d} ) - e_0\mathbf{w}_d,\\
\mathbb{E}_{\mathbf{\beta}}\Big[\ln q(\mathbf{\beta})\Big] & = \frac{1}{2} \sum^C  \ln det(\lambda_{\beta_i} ),\\
\mathbb{E}_{\Delta}\Big[\ln q(\Delta)\Big] & = \frac{1}{2} \ln det(\lambda_{\Delta} ),\\
\mathbb{E}_{\sigma}\Big[\ln q(\sigma)\Big] & = a_n -\ln b_n + \ln \Gamma(a_n)+ (1-a_n)\psi(a_n),\\
\mathbb{E}_{s}\Big[\ln q(s)\Big] & = c_n -\ln d_n + \ln \Gamma(c_n)+ (1-c_n)\psi(c_n),\\
\mathbb{E}_{\mathbf{w}}\Big[\ln q(\mathbf{w})\Big] & = \sum^D e_n - \ln \mathbf{f}_{n_d}+ \ln \Gamma(e_n)+(1-e_n)\psi(e_n),\\
\end{align*}

where $\psi(\cdot)$ is the digamma function. During the optimization, the bound is maximized and the optimization stopped when it reaches a plateau
$|\mathcal{L}(q_n) - \mathcal{L}(q_{n+1})| < \epsilon $.

\subsection{Predictive density}

After we obtained the approximated posterior distribution, we might want to make predictions for new target variables $y_{*}$ based on new observations $\mathbf{x}_{*}$.
Therefore, we require the posterior predictive distribution, which allows us to estimate confidence intervals. 
The posterior predictive distribution for the new target variable of the newly observed sample is calculated by marginalizing the distribution of the target variable given the observation and parameters over the posterior distribution of the parameters. We substitute the posterior distribution for the variational posterior.

\begin{align*}
p(y_{*} | \mathbf{x}_{*}, \mathbf{D}) & = \int\int p(y_{*} \mid \mathbf{x}_{*},\beta,\sigma ) p(\beta,\sigma \mid \mathbf{D})\;d\beta\;d\sigma\\  
 & \approx \int\int p(y_{*} \mid \mathbf{x}_{*},\beta,\sigma ) q(\beta,\sigma )\;d\beta\;d\sigma\\
 & = \int\int  \mathcal{N} \left (y_{*} \mid \mathbf{x}_{*}\beta, \sigma^{-1} \right ) \mathcal{N} \left (\beta \mid \beta_i, \lambda_{\beta_i} \right )\operatorname{Gamma}(\sigma\mid a_n, b_n) \;d\beta\;d\sigma\\
 & = \int  \mathcal{N} \left (y_{*} \mid \mathbf{x}_{*}\beta_i, \sigma^{-1} + \mathbf{x}_{*}^T\lambda_\beta \mathbf{x}_{*}  \right )\operatorname{Gamma}(\sigma\mid a_n, b_n) \;d\sigma\\
 & = St\left( y_{*}, \frac{a_n}{b_n} + \mathbf{x}_{*}^T\lambda_{\beta_i} \mathbf{x}_{*} ,2a_n   \right)
\end{align*}

To obtain the result, we used results for the convolution of normal and Gamma distributions~\cite{Bishop:2006:PRM:1162264, Murphy2013Machine}.
First, we convoluted the two normal distributions to integrate out $\beta$. The marginal distribution of the resulting normal with the Gamma distribution results into a Student's t distribution with mean $\mathbf{x}_{*}\beta_i$, precision $\frac{a_n}{b_n} + \mathbf{x}_{*}^T\lambda_\beta \mathbf{x}_{*}$, and  $2a_n$ degrees of freedom.
The result shows that the predicted uncertainty is the sum of the noise $\sigma^{-1}$ and the variance in client individual weights $\beta_i$. The number of degrees of freedom is approximately the number of observed samples. After we have observed 30 samples we could approximate this with a normal distribution. Interestingly, these samples do not have to come from an individual since it is the number of the overall observed samples. We can also make a prediction for clients that we have never observed before by using the hierarchical prior $\Delta$ and the variance $s$.

\section{Data Analysis}

For a demonstration of the algorithm, we chose a freely available dataset from the machine learning repository (\url{http://archive.ics.uci.edu/ml/datasets/Turkiye+Student+Evaluation/}). We chose the Turkiye Student Evaluation Data Set~\cite{GunduzFokoue:2013}, which is a data set that consists of the evaluation score of courses from the Gazi University in Ankara (Turkey). To rate the individual courses, 28-course specific questions were inquired and 5 additional attributes. An overview of the questions and items is shown in Table~\ref{tab:data}

\begin{table}[H]
\tiny
\begin{center}
  \begin{tabular}{ l | l  }
    \hline
    Item & Description \\
    \hline
 instr & Instructor's identifier \\
class& Course code (descriptor)\\
repeat& Number of times the student is taking this course\\ 
attendance& Code of the level of attendance\\
difficulty& Level of difficulty of the course as perceived by the student; values taken from \\
Q1& The semester course content, teaching method and evaluation system were provided at the start.\\
Q2& The course aims and objectives were clearly stated at the beginning of the period.\\
Q3& The course was worth the amount of credit assigned to it.\\
Q4& The course was taught according to the syllabus announced on the first day of class.\\
Q5& The class discussions, homework assignments, applications and studies were satisfactory.\\
Q6& The textbook and other courses resources were sufficient and up to date.\\
Q7& The course allowed field work, applications, laboratory, discussion and other studies.\\
Q8& The quizzes, assignments, projects and exams contributed to helping the learning.\\
Q9& I greatly enjoyed the class and was eager to actively participate during the lectures.\\
Q10& My initial expectations about the course were met at the end of the period or year.\\
Q11& The course was relevant and beneficial to my professional development.\\
Q12& The course helped me look at life and the world with a new perspective.\\
Q13& The Instructor's knowledge was relevant and up to date.\\
Q14& The Instructor came prepared for classes.\\
Q15& The Instructor taught in accordance with the announced lesson plan.\\
Q16& The Instructor was committed to the course and was understandable.\\
Q17& The Instructor arrived on time for classes.\\
Q18& The Instructor has a smooth and easy to follow delivery/speech.\\
Q19& The Instructor made effective use of class hours.\\
Q20& The Instructor explained the course and was eager to be helpful to students.\\
Q21& The Instructor demonstrated a positive approach to students.\\
Q22& The Instructor was open and respectful of the views of students about the course.\\
Q23& The Instructor encouraged participation in the course.\\
Q24& The Instructor gave relevant homework assignments/projects, and helped/guided students.\\
Q25& The Instructor responded to questions about the course inside and outside of the course.\\
Q26& The Instructor's evaluation system (midterm, assignments, etc.) effectively measured the course objectives.\\
Q27& The Instructor provided solutions to exams and discussed them with students.\\
Q28& The Instructor treated all students in a right and objective manner.\\
\hline
\multicolumn{2}{ l }{Q1-Q28 are all Likert-type, meaning that the values are taken from {1,2,3,4,5}} \\
%Q1-Q28 are all Likert-type, meaning that the values are taken from {1,2,3,4,5}
\hline
  \end{tabular}
\caption{Items and questions in the data set}
\label{tab:data}
\end{center}
\end{table}

The data has a hierarchical structure in a way that each class is measured multiple times. However, we can also see that each instructor has multiple classes, but for modeling this structure we would have to extend the model. Therefore, we neglect this variable for now. The question of interest is, what are the characteristics of a difficult class, and can we predict the difficulty given the questions and items. Since the data in the data set is ordinal, it is clearly not perfectly suited for analysis with the developed algorithm, but it should help in demonstrating the algorithm.
We start the analysis with a linear regression model and the results are shown in Table~\ref{tab:reg}.
% latex table generated in R 3.5.1 by xtable 1.8-3 package
% Sat Oct 27 20:58:34 2018
\begin{table}[ht]
\tiny
\centering
\begin{tabular}{rrrrr|rrrrr}
  \hline
 Item& Estimate & Std. Error & t value & Pr($>$$|$t$|$) & Item& Estimate & Std. Error & t value & Pr($>$$|$t$|$) \\ 
  \hline
nb.repeat & 0.858 & 0.024 & 35.32 & <0.000  &Q14 & 0.082 & 0.048 & 1.72 & 0.086 \\ 
  attendance & 0.453 & 0.011 & 39.58 & <0.000  & Q15 & 0.020 & 0.044 & 0.47 & 0.642 \\ 
  Q1 & 0.065 & 0.027 & 2.36 & 0.018  & Q16 & -0.151 & 0.040 & -3.74 & <0.001 \\ 
  Q2 & 0.035 & 0.035 & 0.99 & 0.320  & Q17 & 0.201 & 0.035 & 5.81 & <0.000 \\ 
  Q3 & -0.020 & 0.031 & -0.65 & 0.517  & Q18 & -0.074 & 0.039 & -1.89 & 0.059 \\ 
  Q4 & 0.026 & 0.033 & 0.80 & 0.424  & Q19 & -0.012 & 0.040 & -0.30 & 0.763 \\  
  Q5 & 0.065 & 0.037 & 1.73 & 0.083  & Q20 & 0.001 & 0.042 & 0.02 & 0.987 \\ 
  Q6 & -0.019 & 0.034 & -0.56 & 0.575  & Q21 & 0.024 & 0.046 & 0.52 & 0.605 \\ 
  Q7 & 0.020 & 0.038 & 0.52 & 0.603  & Q22 & 0.048 & 0.046 & 1.03 & 0.302 \\ 
  Q8 & 0.064 & 0.036 & 1.80 & 0.073  & Q23 & -0.033 & 0.043 & -0.75 & 0.453 \\  
  Q9 & -0.037 & 0.030 & -1.23 & 0.219  & Q24 & 0.053 & 0.039 & 1.34 & 0.181 \\ 
  Q10 & -0.091 & 0.041 & -2.21 & 0.027  & Q25 & 0.087 & 0.042 & 2.11 & 0.035 \\  
  Q11 & -0.028 & 0.031 & -0.91 & 0.363  & Q26 & -0.084 & 0.036 & -2.32 & 0.020 \\  
  Q12 & -0.026 & 0.030 & -0.85 & 0.396  & Q27 & 0.032 & 0.032 & 1.02 & 0.308 \\  
  Q13 & 0.013 & 0.042 & 0.30 & 0.767  & Q28 & 0.003 & 0.036 & 0.09 & 0.927 \\  
   \hline
\end{tabular}
\caption{Result of the linear regression to explain the difficulty of a class}
\label{tab:reg}
\end{table}

The regression results show that both items, the number of a repeated taking of the class and attendance indicate a class with higher difficulty. Apparently, students are more inclined to visit the lectures, when the content is difficult. Regarding the questions, the 1st, 10th, 16th, 17th, 25th, and 26th show a significant relationship to the course difficulty. The first question describes if the course topic and content were presented adequately at the start of the course, potentially instructors of difficult classes tend to state the goals and evaluation method of difficult classes more clearly in the beginning. The 10th item is negatively related to the difficulty, which states if the expectancies of the course result were met. Supposedly, students are more pleased with the outcome of the class if its perceived difficulty is lower.  
The 16th question is negatively associated with the difficulty, which asks if the instructor was committed to the course and easily understandable. Enthusiastic and clearly understandable instructors are associated with easier overall classes. Interestingly, the 17th question inquires if the instructor arrived in time and is positively associated with the class difficulty. This suggests that if the instructor arrives early to the class, its content might be more difficult.
Similarly, the 25th question indicates that if the instructor responded to questions inside and outside the class, the course was more difficult.
Question 26 indicates that the evaluations methods of courses with higher difficulty were perceived as effective to measure the course objective.

\begin{figure}[H]%
    \centering
   \includegraphics[width=0.8\textwidth]{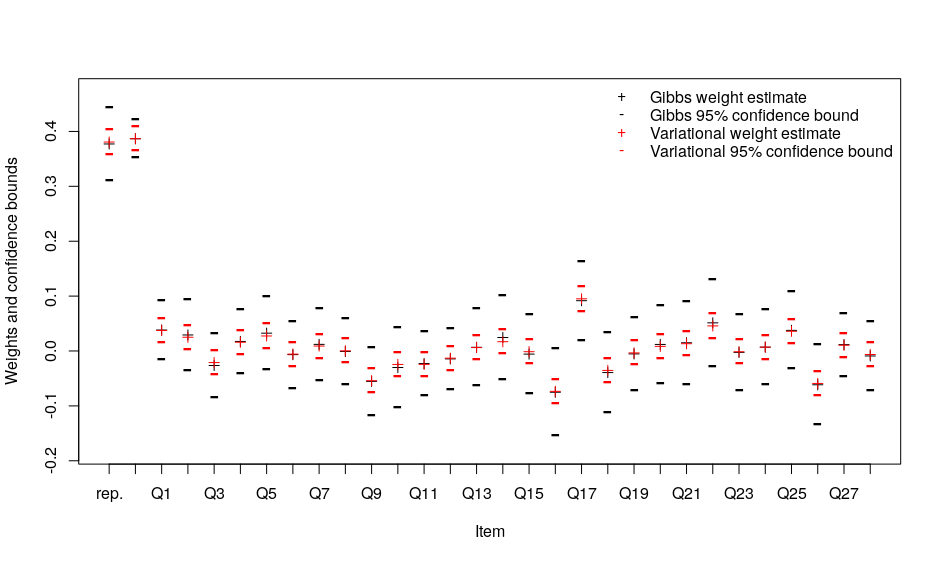}
    \caption{Comparison of the estimated weights and confidence interval of the variational approximation and a Gibbs sampler}%
    \label{fig:comp}%
\end{figure}

This provides some overview of the data. We could fit the variational model to the data to estimate the weights and confidence intervals. However, keep in mind that variational inference only provides an approximation and will underestimate the variance in the posterior distribution. To visualize this, we train the model on the data and use a Gibbs sampler for the estimation of the same model. The weight estimates and 95\% confidence intervals are visualized in Figure~\ref{fig:comp}

The variational model also provides a ranking of the importance of the features, which we compare to the feature ranking of the Lasso regression\cite{Tibshirani1994}.
For the estimation of the ranking for the Lasso regression, we add the class variable and rank the features according to its first appearance in the lambda path.
The hierarchical model, however, does not have a variance parameter $\mathbf{w}_i$ for each class. Therefore, we use the estimated weights for their ranking. Because we used two different estimates for ranking of the features for the variational model, we present the results in separately in Table ~\ref{tab:ranking}.

\begin{table}[!htb]
\tiny
    \begin{minipage}{.5\linewidth}
      \centering
    %    \begin{tabular}{ll}
    %        1 & 2
    %    \end{tabular}

%\begin{table}[ht]
%\centering
\begin{tabular}{rll}
  \hline
 & Lasso regression& Variational model \\ 
  \hline
1 & attendance & attendance \\ 
  2 & nb.repeat & nb.repeat \\ 
  3 & Q17& Q17 \\ 
  4 &Q9  & Q16 \\ 
  5 &Q11  & Q26 \\ 
  6 &Q12  & Q9 \\ 
  7 &Q10  & Q22 \\ 
  8 &Q16  & Q1 \\ 
  9 & Q1 & Q25 \\ 
  10 & Q22 & Q18 \\ 
  11 & Q25 & Q5 \\ 
  12 & Q26 & Q2 \\ 
  13 & Q18 & Q10 \\ 
  14 &Q4  & Q11 \\ 
  15 & Q2 & Q3 \\ 
  16 & Q5 & Q14 \\ 
  17 & Q14 & Q4 \\ 
  18 & Q3 & Q21 \\ 
  19 &Q21  & Q12 \\ 
  20 & Q24 & Q27 \\ 
  21 & Q7 & Q7 \\ 
  22 & Q27 & Q20 \\ 
  23 &Q13  & Q24 \\ 
  24 & Q19 & Q28 \\ 
  25 &Q23  & Q13 \\ 
  26 &Q28  & Q6 \\ 
  27 & Q20 & Q19 \\ 
  28 &Q6  & Q23 \\ 
  29 &Q8  & Q15 \\ 
  30 & Q15 & Q8 \\ 
   \hline
\end{tabular}
      \caption*{Comparison of the ranking for the questions}

    \end{minipage}%
    \begin{minipage}{.5\linewidth}
      \centering
\begin{tabular}{rll}
  \hline
 & Lasso Regression& Variational model \\ 
  \hline
  1 & class 7 & class 7 \\ 
  2 & class 13 & class 8 \\ 
  3 & class 2 & class 2 \\ 
  4 & class 1 & class 11 \\ 
  5 & class 6 & class 4 \\ 
  6 & class 11 & class 6 \\ 
  7 & class 10 & class 3 \\ 
  8 & class 5 & class 12 \\ 
  9 & class 3 & class 9 \\ 
  10 & class 9 & class 5 \\ 
  11 & class 4 & class 10 \\ 
  12 & class 12 & class 1 \\ 
  13 & class 8 & class 13 \\ 
   \hline
\end{tabular}
        \caption*{Comparison of the ranking for the classes}

    \end{minipage} 
    \caption{Comparison of estimated feature importance for Lasso regression and the variational model}
\label{tab:ranking}
\end{table}

The importance ranking of the questions is similar in both models. However, the ranking for the classes, where we use the absolute value of the weights for the variational model appears to differ. Although both models agree on the importance of class 7, the Lasso regression ranks class 13 as second most important whereas the variational model ranks this last. Utilizing the individual classes weights for ranking might not be a reliable method to rank the importance.

As a next objective, we are interested in the prediction performance of the variational model in comparison to similar models.
Therefore, we use 10 fold cross where we train the model on 9 folds and predict the remaining fold. Before we estimate the mean square error of that fold, we round the predictions. For the estimation of the cross-validated error, we use linear regression, lasso regression, ridge regression, the variational hierarchical Bayesian regression, and a linear mixed-effects model with an individual intercept for each class.

\begin{table}[ht]
\small
\centering
\begin{tabular}{r|ccccc}
  \hline
            & Regression & Lasso  &  Variational model& LME model&Ridge\\ 
  \hline
Mean square error &1.45 &  1.43   & 1.45 & 1.45 & 1.54\\
   \hline
\end{tabular}
\caption{Cross validated mean square error of the various models }
\label{tab:predResults}
\end{table}

The results of the cross-validated error in Table~\ref{tab:predResults} show that the Lasso regression has the lowest prediction error followed by the linear regression. The error of the hierarchical regression is higher but still lower than the prediction error of the ridge regression. Its prediction error is comparable to the error of the linear mixed effect model. Apparently, the setup of the analysis and data are not favoring the hierarchical models, however, we hope that this provides some intuition into the models use, capabilities, and limitations.

\section{Concluding Remarks} %Discussion and Concluding Remarks
 
We derived a variational approximation for a hierarchical Bayesian regression model and demonstrated its application on real-world data set.
The main benefit of this model might be that it can be estimated considerably faster than the same model using Gibbs sampling. Therefore, we can see its use for application where the model has to be estimated quickly, repeatedly, or on a bigger dataset. Furthermore, we hope that his document could shed some light and raise interest in Bayesian modeling, variational approximation, and its potential use in applications and data analysis.

\section{Acknowledgements}

We want to thank the machine learning repository and the publisher of the data because they made this analysis possible.

\newpage

\bibliographystyle{plain}

\bibliography{references.bib}

\end{document}